\documentclass[letter,twocolumn]{jpsj3}

\usepackage{txfonts}
\usepackage{amsmath,amssymb} 
\usepackage{graphicx}
\usepackage{color}
\usepackage{bm}
\usepackage{mathrsfs}
\usepackage{setspace}

\DeclareMathOperator{\diag}{diag}

\newcommand{\FRAC}[2]{{\textstyle \frac{#1}{#2}}}

\newcommand{\msm}[1]{\mspace{-#1mu}}
\newcommand{\msp}[1]{\mspace{#1mu}}
\newcommand{\MINUS}{\msm2 - \msm2}
\newcommand{\PLUS}{\msm2 + \msm2}
\newcommand{\Eq}{\msm3 = \msm3}
\newcommand{\Equiv}{\msm2 \equiv \msm2}

\newcommand{\supS}{^{\, {\scriptscriptstyle S}}}
\newcommand{\supL}{^{\, {\scriptscriptstyle L}}}
\newcommand{\subS}{_{\msm2 {\scriptscriptstyle S}}}
\newcommand{\subL}{_{\msm2 {\scriptscriptstyle L}}}
\newcommand{\ssSUP}[1]{^{\scalebox{0.6}{#1}}}
\newcommand{\ssSUB}[1]{_{\scalebox{0.6}{#1}}}

\newcommand{\baa}{\bm{a}}
\newcommand{\bdd}{\bm{d}}
\newcommand{\bee}{\bm{e}}
\newcommand{\bkk}{\bm{k}}
\newcommand{\bpp}{\bm{p}}
\newcommand{\brr}{\bm{r}}
\newcommand{\btt}{\bm{t}}
\newcommand{\bzero}{{\bm{0}}}
\newcommand{\bbeta}{\bm{\eta}}
\newcommand{\bzeta}{\bm{\zeta}}
\newcommand{\bLL}{\bm{L}}

\newcommand{\sfA}{\mathsf{A}}
\newcommand{\sfC}{\mathsf{C}}
\newcommand{\sfD}{\mathsf{D}}
\newcommand{\sfF}{\mathsf{F}}
\newcommand{\sfL}{\mathsf{L}}
\newcommand{\sfv}{\mathsf{v}}
\newcommand{\sfO}{\mathsf{O}}
\newcommand{\sfR}{\mathsf{R}}
\newcommand{\sfS}[1]{\mathsf{S}_{#1}}

\newcommand{\dm}{\msm3 \circ \msm3}

\newcommand{\ulD}{\underline{D}}
\newcommand{\ulP}{\underline{\varPi}}
\newcommand{\ulbP}{\underline{\bm{\varPi}}}

\newcommand{\sE}{\mathscr{E}}

\title{Chiral Phonons in a Cubic Lattice}

\author{Hirokazu Tsunetsugu$^1$ and Hiroaki Kusunose$^{2}$}
\inst{$^1$The Institute for Solid State Physics, 
The University of Tokyo, 
Kashiwanoha 5-1-5, Chiba 277-8581, Japan \\
$^2$Department of Physics, Meiji University, Kawasaki 214-8571, Japan} 

\abst{
We have developed a theory for the energy dispersion 
of chiral phonons in a simplest cubic lattice. 
Among all the phonon modes, only the optical triplet modes 
exhibit the intrinsic characteristics of chiral phonons 
near $\bm{k}=\bm{0}$, and 
we examine their energy splitting in detail by 
analyzing the dynamical matrix.  
To first order in $\bkk$,  
the splitting is described by a spin-1 Weyl Hamiltonian, 
and helicity becomes a good quantum number.  
It asymptotically coincides with the crystal angular momentum 
for $\bkk \parallel \langle 1 1 1 \rangle$ 
up to a global sign.  
The Hamiltonian incorporates a pseudoscalar coupling constant 
associated with electric toroidal monopoles,  
determined by the spatial configuration of the stiffness tensors.  
}

\begin{document}
\maketitle

Phonons with nonzero angular momentum have 
recently attracted significant attention~\cite{WangTT,Juraschek2025}.
Similar to circularly polarized photons, 
they carry intrinsic angular 
momentum~\cite{Zhang2014,Zhang2015,Hamada2018,Ohe2024,Zhang2025} 
and are expected to generate significant magnetic moments 
through (resonant) interactions with electron 
spins~\cite{Hamada2020,Zhu2018,Luo2023,Shabala2024,%
Ren2024,Chaudhary2024,Wang2024}.
The concept of phonon \textit{spin}, 
first proposed in the 1960s~\cite{Vonsovskii,Levine,Teuchert1974,Bozovic1984}, 
has recently been revitalized~\cite{Zhang2022,HT,Kato2023,Tateishi2025}, 
particular in chiral materials~\cite{Bousquet2025} 
lacking spatial inversion and/or mirror symmetries, 
where phonon energy dispersions split according to 
their angular momentum~\cite{Pine1969,Ishito2022,Ishito2023,Ueda2023,Oishi2024}.
When the phonon's angular momentum is parallel to 
its propagating direction, 
its symmetry property is genuinely chiral, 
and such modes are termed \textit{chiral phonons}.  
In contrast, phonons possessing intrinsic angular momentum 
that is not necessarily parallel to the propagation direction 
are more recently referred to as \textit{axial phonons}~\cite{Juraschek2025}.

The properties of axial phonons have been 
mainly investigated in monoaxial crystals, 
such as trigonal or hexagonal systems.
In these crystals, phonon eigenmodes are 
characterized by right- and left-handed rotational 
modes which split from a two-dimensional irrep.
When the phonon propagates along the screw-axis direction, 
the crystal angular momentum (CAM) becomes a good quantum 
number\cite{Zhang2014,Zhang2015,Bozovic1984,Tatsumi2018,Zhang2022,HT,Kato2023}.
Furthermore, optical modes near the $\Gamma$-point 
in the Brillouin zone exhibit linearly split dispersions.
By contrast, cubic crystals possess multiple equivalent 
screw axes and support three-dimensional irreps, 
necessitating a fundamentally different classification 
of chiral phonons from that in monoaxial systems.

\begin{figure}[b]
\centering{
\includegraphics[height=4cm]{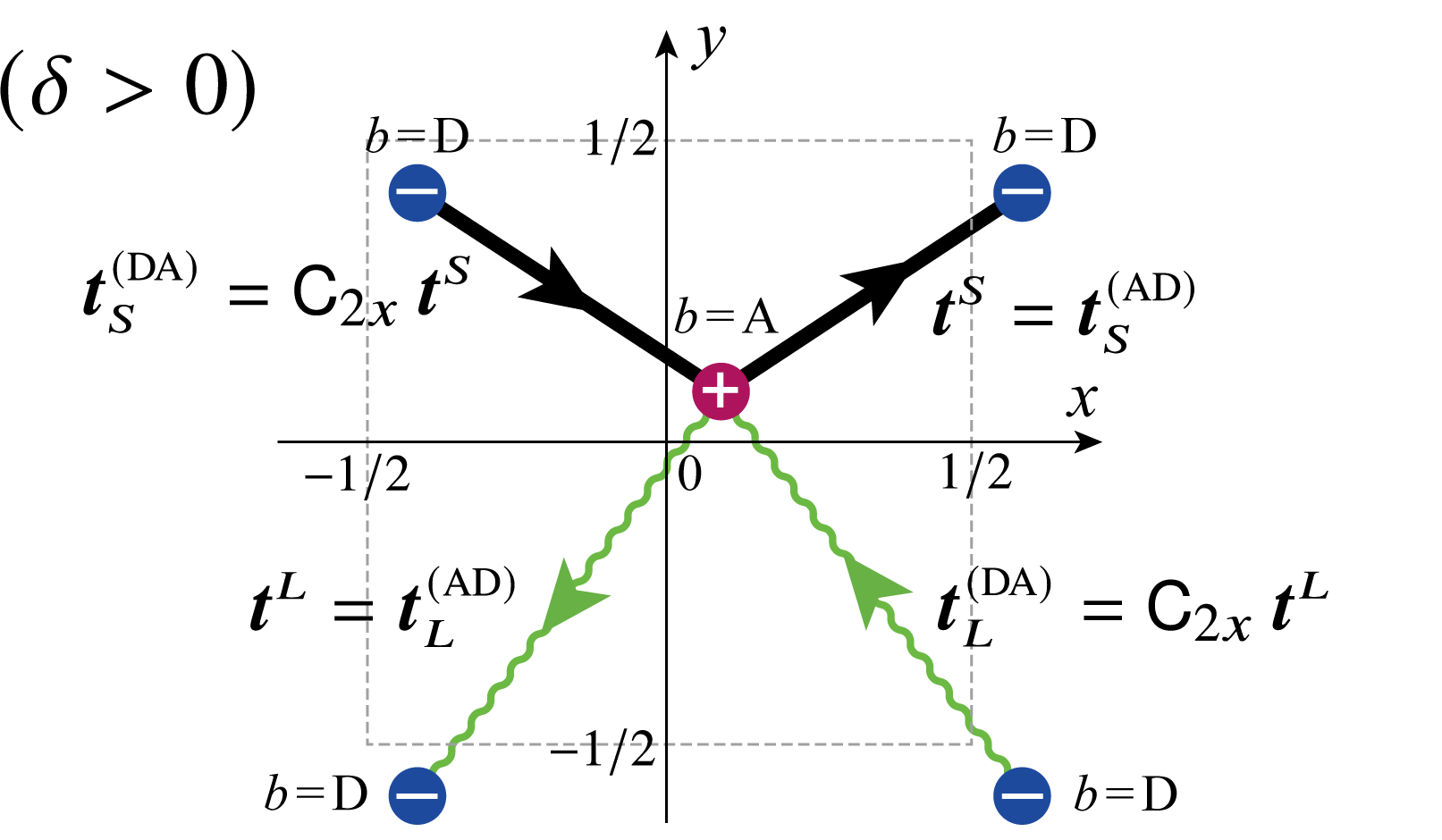}
}
\caption{
(Color online) 
Four bonds connecting the sublattices A and D. 
Short and long bonds are drawn with straight and wavy lines. 
The symbol $\pm$ shows the sign of the $z$-coordinate of each 
atomic position.  
Note $\btt \supL \Eq - \btt \supS$ when $\delta \Eq 0$.  
}
\label{fig1}
\vspace{-10pt}
\end{figure}

In a previous letter, we have studied energy dispersion of 
phonons in a chiral crystal with one screw axis~\cite{HT,Kato2023}.  
We have derived an analytical formula of energy split 
depending CAM 
represented in terms of microscopic model parameters 
of stiffness constants.  
This is the simplest example of chiral crystals, and 
other chiral crystals have more complex structures.  
In this letter, we will study the same problem for the chiral 
systems with the highest lattice symmetry, i.e., cubic symmetry.  
Cubic chiral crystals usually contain many atoms 
in their unit cell and this makes analysis tedious.  
Thus, we will study a simple toy lattice 
with the minimal number of sublattices in order to 
find general properties specific to chirality of the lattice.  
The results obtained from studying it 
should be helpful for analyzing complicated 
phonon spectra in real chiral materials with cubic symmetry.  

The chiral lattice to be studied is a distorted version of 
face centered cubic (Dfcc) lattice with the space group symmetry 
$P2_1 3$ ($T^4$, No.~198).  
Its unit cell is a cube containing four atoms of an identical 
element occupying the 4a Wyckoff positions.  
Let us use the units that the lattice constant is 1, and then 
the lattice vectors are a triad of unit vectors 
$\{ \bee _1, \msm2 
\bee _2, \msm2 
\bee _3 \}$  
such as $\bee _1 = (1,0,0) = \bee _x$ \textit{et al.}, 
and 
we also define $\bbeta _0 = (1,1,1)$.  
As for the indices for Cartesian components, we will use the notation 
(1,2,3) and $(x,y,z)$ interchangeably throughout this letter.  
Atomic positions are shifted as 
\begin{equation}
\brr \ssSUB{A} \Eq \delta \bbeta _0 , \ \  
\brr \ssSUB{B} \Eq \baa _1 + \delta \bbeta _2 , \ \  
\brr \ssSUB{C} \Eq \baa _2 + \delta \bbeta _3 , \ \  
\brr \ssSUB{D} \Eq \baa _3 + \delta \bbeta _1 , 
\label{eq:sublattice}
\end{equation}
where 
$\delta$ is a dimensionless pseudoscalar parameterizing
the chiral distortion.  
Note $\baa _i = \FRAC12 (\bbeta _0 - \bee _i)$ is 
the original position in the regular fcc lattice.
The vectors $\bbeta _i$'s are 
$\bbeta _i = \sfC _{2 i} \msp2 \bbeta _0$, 
where 
$\sfC _{2 i}$ 
is the C$_2$ rotation operator 
about the 
$\bee _i$-axis \textit{e.g.}, 
$\sfC _{21} = \sfC _{2x} \Eq \diag \msp1 (1,-1,-1)$.  
We let 
$\sfS{b}$ ($b \msm2 \in \msm2 \{ \mbox{A, B, C, D} \}$)
denote the C$_3$ rotation 
about the axis $\bbeta _i$
appearing in the definition \eqref{eq:sublattice} 
for the sublattice $b$.  
We also use the shorthand notation for orthogonal 
transformations 
$\sfO \msp2 \dm \msp2 \sfA \Equiv \sfO \msp2 \sfA \msp2 {}^t \sfO$, 
where ${}^t \sfO$ represents transposed matrix.  
In the regular fcc lattice, its cubic unit cell 
contains 24 nearest-neighbor bonds 
of equal length.  
When $\delta \ne 0$, half of these bonds shorten while 
the other half lengthen.  
Their lengths are given by 
$l_{\mp} \msm3 = \msm3 
[1/2 \msm2 + \msm2 8\delta^2 \msm2 \mp \msm2 2\delta ]^{1/2}$.  

We build a minimal model of phonons for this chiral lattice 
that continuously reaches a 
nonchiral limit without changing unit cell.  
In this nonchiral limit ($\delta =0$), 
we devise a lattice recovering 
the inversion symmetry such that its space group is 
$Pa\bar{3}$ ($T_h^6$, No.205), 
which is a minimal supergroup of $P2_1 3$.  
In all the cases, we take account of all 
of 24 nearest neighbor bonds in the unit cell.  
They are labeled by a sublattice pair 
$(b,b')$ 
of the end atoms 
together with the index 
$p \msm2 \in \msm2\{S, L \}$ indicating bond length.  
For each bond, we also assign 
an end-to-end vector $\btt _p^{(bb')}$ 
and 
an stiffness tensor $\sfv_p ^{(bb')}$.  
This tensor defines the bond's 
deformation energy as 
$\FRAC12 (\bdd - \bdd ') \cdot \sfv _p^{(bb')} (\bdd - \bdd ') $ 
with the displacement vectors of the two end atoms,  
$\bdd = \bdd _b (\brr )$ and 
$\bdd ' = \bdd _{b'} (\brr + \btt _p^{(bb')} )$.   

It suffices to define one representative for each of 
short and long bonds, and we choose 
$\btt \supS \Eq \btt \subS \ssSUP{(AD)} = 
\brr \ssSUB{D} \MINUS \brr \ssSUB{A}$ and 
$\btt \supL \Eq \btt \subL \ssSUP{(AD)} 
\Eq \btt \supS \MINUS 2\baa _3 
\Eq -\btt \supS \MINUS 8 \delta \baa _1$.  
See Fig.~\ref{fig1}.  
Correspondingly, the representative tensors are 
$\sfv \supS \Eq \sfv \subS \ssSUP{(AD)}$ and 
$\sfv \supL \Eq \sfv \subL \ssSUP{(AD)}$.   
The others are given by applying symmetry operations on them: 
$\sfv _p^{(bb')} = \sfR ^{(bb')} \circ \sfv ^p$.  
Note that for each of twelve $(bb')$ pairs, 
there exists 
a unique symmetry operation $\sfR ^{(aa')}$ 
in the T-point group 
which satisfies 
$\btt _p^{(bb')} = \sfR ^{(bb')} \btt ^p$: 
the reference is $\sfR \ssSUP{(AD)} $=$ \mathsf{1}$, 
the C$_2$ rotations are 
$\sfR \ssSUP{(DA)} \msm2 = \msm2 \mathsf{C}_{2x}$, 
$\sfR \ssSUP{(BC)} \msm2 = \msm2 \mathsf{C}_{2y}$, 
$\sfR \ssSUP{(CB)} \msm2 = \msm2 \mathsf{C}_{2z}$, 
and the others are C$_3$ rotations
$\sfR \ssSUP{(AB)} \msm2 = \msm2 {}^t \sfR \ssSUP{(AC)} 
                   \msm2 = \msm2 {\mathsf{S}\ssSUB{A}}$, 
$\sfR \ssSUP{(BC)} \msm2 = \msm2 {}^t \sfR \ssSUP{(BD)} 
                   \msm2 = \msm2 {\mathsf{S}\ssSUB{D}}$, 
$\sfR \ssSUP{(CB)} \msm2 = \msm2 {}^t \sfR \ssSUP{(CA)} 
                   \msm2 = \msm2 {\mathsf{S}\ssSUB{B}}$, 
$\sfR \ssSUP{(BA)} \msm2 = \msm2 {}^t \sfR \ssSUP{(DB)} 
                   \msm2 = \msm2 {\mathsf{S}\ssSUB{C}}$.
Since the bond center position has no symmetry, 
$\sfv \supS$ and $\sfv \supL$ are unconstrained 
as far as they are symmetric and positive definite, 
and each tensor is characterized by 6 parameters.  
The lattice with the distortion parameter $-\delta$ 
is an inversion image of the original with $+\delta$, 
and 
this leads to the relation 
$\sfv \supS (\delta) = \sfv \supL (-\delta) $.  
This immediately shows the symmetry in 
the nonchiral limit  
$\sfv \supS (0) \Eq \sfv \supL (0) 
\msm3 = : \msm2 \bar{\sfv}$, 
but none of $\bar{\sfv}$'s elements needs to vanish. 
This is because the bond center positions have 
no symmetry even when $\delta = 0$, and in this sense 
the nonchiral limit differs from the fcc lattice,  
in which 
the higher $Fm\bar{3}m$ symmetry 
ensures the presence of three mirror symmetry operations 
at each bond center.

\begin{figure*}[t]
\centering{
\includegraphics[width=0.9\textwidth]{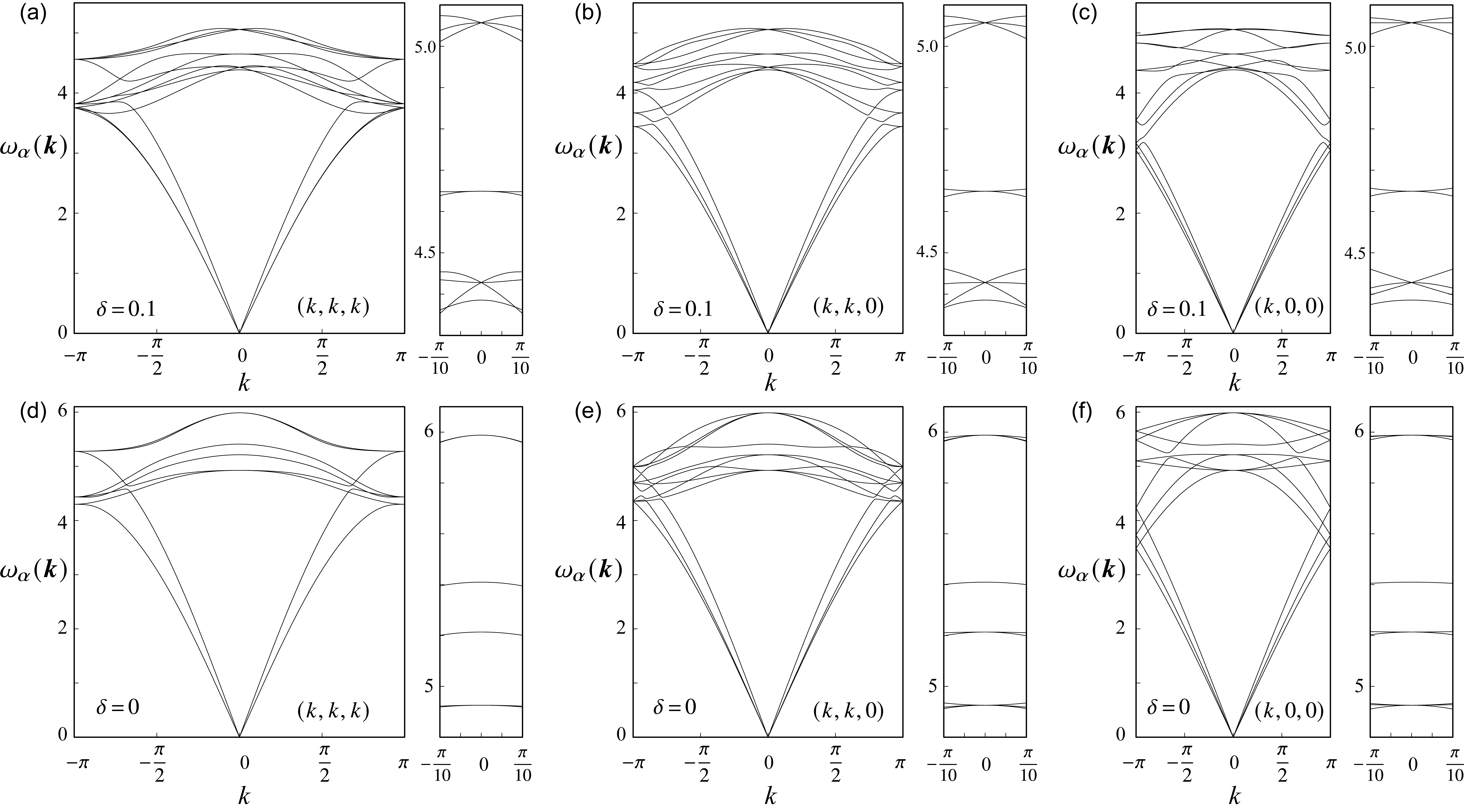}
}
\caption{
Phonon energy dispersion in the Dfcc lattice along 
high-symmetry axes in the Brillouin zone with 
a magnified view near $\bkk =0$.  
(a)-(c) 
Results for a chiral lattice with $\delta = 0.1$.  
Stiffness tensor elements 
$\{ v_{\mu \nu}\} $=
$(v_{xx}, v_{yy}, v_{zz}, v_{yz}, v_{zx}, v_{xy})$ are 
$(2.151, \, 2,103, \, 1.246, \, -0.462, \, -0.467 , \, 0.627)$ 
for $\sfv \supS$ and 
$(0.894, \, 1.219, \, 0.638, \, 0.264, \, 0.028, \, 0.340)$ 
for $\sfv \supL$.  
Energy levels at $\bkk \! = \! \bm{0}$ are 
T$^{\, 0}$, A, T$^{\, -}$, E, T$^{\, +}$ in ascending order.  
(d)-(f) 
Results for a non-chiral lattice ($\delta = 0$) 
with $\{ v_{\mu \nu}\}$=
$(2.015, \, 2.462, \, 1.023, \, 0.087, \, -0.064, \, 0.727)$ for 
both of $\sfv \supS$ and $\sfv \supL$.  
Order of the energy levels at $\bkk =\bm{0}$ is now 
T$^{\, 0}$, T$^{\, -}$, E, A, T$^{\, +}$.  
}
\label{fig:disp}
\end{figure*}

In Fourier space, the lattice deformation energy is written 
in terms of the 12-dimensional displacement vector 
$\vec{d}(\bkk) \Eq 
{}^t (\bdd \ssSUB{A} (\bkk), \cdots , \bdd \ssSUB{D} (\bkk) )$ 
as 
$\sE \Eq
\sum_{\bkk} \vec{d}(\bkk )^\dagger 
\ulD (\bkk) \, \vec{d} (\bkk) $.  
Its coefficient is the dynamical matrix $\ulD (\bkk ) 
$ 
made of 16 sublattice blocks 
$\sfD  ^{(bb')} (\bkk ) = 
 \sfD \subS ^{(bb')} (\bkk ) +\sfD \subL ^{(bb')} (\bkk ) 
$, 
and each block is a $3 \times 3$ matrix.  
The off-diagonal blocks are represented as 
\begin{equation}
  \sfD _p^{(bb')}(\bkk) = 
- \sfv _p^{(bb')} \exp \msp1 \bigl(  i \bkk \cdot \btt_p^{(bb')} \bigr)
- \sfv _p^{(b'b)} \exp \msp1 \bigl( -i \bkk \cdot \btt_p^{(b'b)} \bigr) . 
\end{equation}

The diagonal blocks are $\bkk$-independent and 
given as 
$ \sfD _p ^{(bb)} \Eq
- \sum_{b' \ne b} \sfD _p ^{(bb')} (\bzero )$.  
We solve the eigenvalue problem 
$\ulD (\bkk ) \msp2 \vec{d}_{\alpha} (\bkk) 
\msm2 = \msm2 
\lambda _{\alpha} (\bkk) \msp2 \vec{d}_{\alpha} (\bkk)$, 
and then the phonon energies are 
obtained as 
$\omega_{\alpha} (\bkk ) \msm2 = \msm2 
\sqrt{\lambda_{\alpha} (\bkk)}$ 
with setting the atomic mass $M_{\mathrm{atom}}=1$.  
The calculated phonon energy dispersion is shown 
in Fig.~\ref{fig:disp}.  

First, let us 
analyze the level structure at $\bkk \Eq \bzero$ 
by diagonalizing 
$\ulD (\bzero)$. 
This matrix has the full symmetry of the point group $T$, 
and each of its eigenvectors $\{ \vec{d}_{\alpha} \}$ belong to one 
of the three irreducible representations (irreps) A, E, and T.  
The two-dimensional irrep E is further decomposed into 
two complex one-dimensional irreps 
$\mathrm{E} = {}^1 \mathrm{E} \oplus {}^2 \mathrm{E}$.  
Bases of ${}^1$E acquire the phase factor 
$\zeta \msm2 \equiv \msm2 e^{2\pi i/3}$ 
upon C$_3$ rotation, while $\zeta^*$ for those of ${}^2$E, 
and their presence is a manifestation of 
chiral lattice structure.  
The sublattice space has dimension four and it is decomposed 
as A$\oplus$T, 
while the displacement vector $\bdd$ should belong 
to T-irrep.  
Their product space $\{ \vec{d} (\bm{0}) \}$ is thus decomposed as 
$(\mathrm{A} \oplus \mathrm{T} ) \otimes \mathrm{T} \Eq 
\mathrm{A} \oplus \mathrm{E} \oplus 3\mathrm{T}$.  
Basis sets of the irreps are obtained by proper symmetrization.  

The A-irrep basis is 
$\vec{e}\msp2 (\mathrm{A}) \Eq 
{}^t (\bbeta _0, \bbeta_2 , \bbeta_3 , \bbeta_1)/\sqrt{12}$, 
and this is nothing but the directions of 
atomic position distortion in Eq.~\eqref{eq:sublattice}.  
The chiral bases of E-irrep are
$\vec{e} \msp2 ({}^1 \mathrm{E}) \Eq 
{}^t (\bzeta _0, \bzeta _2 , \bzeta _3 , \bzeta _1)/\sqrt{12}
\Eq \vec{e} \msp2 ({}^2 \mathrm{E})^* $, 
where $\bzeta _0 \equiv {}^t (1,\zeta,\zeta^*)$ and 
$\bzeta _i^{} \equiv \sfC_{2i} \msp2 \bzeta _0^{}$.  
We choose basis sets of the three T-irreps 
T$^0$, T$^1$, and T$^2$ in the following way
\begin{equation}
\\[-4pt]
\vec{e} \msp2 (\mathrm{T}_{\mu}^{\msp2 n}) \Eq
{}^t \bigl(
\mathsf{S} \ssSUB{A}^n \msp2 \bee _{\mu}, \msp2
\mathsf{S} \ssSUB{B}^n \msp2 \bee _{\mu}, \msp2
\mathsf{S} \ssSUB{C}^n \msp2 \bee _{\mu}, \msp2
\mathsf{S} \ssSUB{D}^n \msp2 \bee _{\mu}  \msp1
\bigr)/2 , 
\ \ \ 
(\mu = x,y,z ).
\label{eq:T-basis}  
\end{equation}
Since the A and ${}^{1,2}$E irreps exist only one set each, 
they are automatically eigenvectors of $\ulD (\bm{0})$. 
Their eigenvalues are $\lambda (\mathrm{A}) \Eq 
4 (v_{xx}^+ \PLUS v_{yy}^+ \PLUS 2 v_{yz}^+)$
and 
$\lambda (\mathrm{E}) \Eq 
4 (v_{xx}^+ \PLUS v_{yy}^+  \MINUS v_{yz}^+)$
where 
$\sfv ^{\pm}  
\Equiv \sfv \supS \pm  \sfv \supL$.  
The T$^0$-mode describes uniform displacements in all the sublattices, 
and thus $\vec{e} \msp2 (\mathrm{T}_{\mu}^0)$'s 
should be zero-mode eigenvectors: 
$\lambda (\mathrm{T}^0) \Eq 0$.  
The only nontrivial part is about T$^{1}$ and T$^{2}$, 
and the problem is reduced to diagonalizing a 2$\times$2 matrix 
$\mathcal{D} \, $: 
$
\langle \vec{e}\msp2 (\mathrm{T}_{\mu}^{n}) \msp2 | \msp2 
 \ulD (\bm{0}) \msp2 | \msp2 
 \vec{e}\msp2 (\mathrm{T}_{\mu'}^{n'}) \rangle \Eq 
 \delta_{\mu \mu'}
 \mathcal{D}_{n n'}
$
with 
$\mathcal{D}_{11} \Eq 4 (v_{xx}^+ \PLUS v_{yy}^+ )$, 
$\mathcal{D}_{22} \Eq 4 (v_{xx}^+ \PLUS v_{zz}^+)$, and 
$\mathcal{D}_{12} \Eq 
 \mathcal{D}_{21} \Eq 4 v_{yz}^+$.
Its diagonalization is immediate and the result is 
$\lambda (\mathrm{T}^{\pm} ) \Eq 
V_0 \pm |\Delta (\mathrm{T})|$ 
with
$V_0 \Equiv 4v_{xx}^+ \PLUS 2v_{yy}^+ \PLUS 2v_{zz}^+$ 
and 
$\Delta (\mathrm{T}) \Equiv 
2 (v_{yy}^+ \MINUS v_{zz}^+ ) + 4 i v_{yz}^+ 
=
|\Delta (\mathrm{T}) | \, e^{i \theta}
$.
Their eigenvectors are 
$\vec{e} \msp2 (\mathrm{T}_{\msm2 \mu}^{+})  
\Eq 
c_{\theta} \msp2 \vec{e} \msp2 (\mathrm{T}_{\msm2 \mu}^1) + 
s_{\theta} \msp2 \vec{e} \msp2 (\mathrm{T}_{\msm2 \mu}^2)$, 
and
$\vec{e} \msp2 (\mathrm{T}_{\msm2 \mu}^{-}) 
\Eq 
-s_{\theta} \msp2 \vec{e} \msp2 (\mathrm{T}_{\msm2 \mu}^1) + 
 c_{\theta} \msp2 \vec{e} \msp2 (\mathrm{T}_{\msm2 \mu}^2)$, 
where 
$c_{\theta} \Eq \cos (\theta /2)$ and 
$s_{\theta} \Eq \sin (\theta /2)$. 

We now discuss the phonon energy split around $\bkk = \bzero$ 
up to linear order in $\bkk$.  
We studied this problem for monoaxial chiral lattices  
in our previous letter~\cite{HT,Kato2023}
by analyzing the dynamical matrix projected 
into a specific CAM subspace. 
When $\bkk$ is parallel to one of 
the trigonal axes $\bbeta$'s, 
we can try a similar analysis, but it is more 
desirable to clarify general behavior for $\bkk$ 
in an arbitrary direction.  
To this end, we follow the idea of the $\bkk \cdot \bpp$ perturbation 
theory and 
expand the full $\ulD (\bkk)$ up to 
linear order of $\bkk$: 
$\ulD (\bkk ) \sim 
 \ulD _0 + \bkk \cdot \ulbP$, and 
$\ulbP \Equiv 
\nabla _{\bkk} \msp2 \ulD (\bkk ) \, \big|_{\bkk = \bzero} 
= (\ulP _x, \ulP _y, \ulP _z)$. 
Note that 
$\ulP _{\mu}$'s 
are hermitian and pure imaginary: 
their sublattice block matrices $\mathsf{\Pi} _{\mu}^{(bb')}$ 
are pure imaginary symmetric, and 
$\mathsf{\Pi} _{\mu}^{(b'b)} \Eq - \mathsf{\Pi} _{\mu}^{(bb')}$.
Thus, diagonal blocks are null, 
$\mathsf{\Pi} ^{(bb)} = \mathsf{0}$.    

The matrix elements of 
$\ulbP$
reduced to each eigenspace 
$\Gamma$ 
determine the linear slope of the eigenvalues, 
$\nabla_{\bkk} \msp2 \lambda_{\msp2 \Gamma} (\bkk ) \, |_{\bkk = \bzero}$. 
We have checked that 
$\ulbP$ 
behaves as a vector under rotations as expected 
and thus belongs to the T-irrep.  
This immediately concludes that the reduced matrix elements
should vanish in the A- and ${}^{1,2}$E-subspaces.  
This means that the level split starts from the order 
higher than linear $\bkk$ for the A- and ${}^{1,2}$E-levels.  
In the T$^0$-subspace, 
$\langle \vec{e} \msp2 (\mathrm{T}_{\mu}^ {\msp2 0} ) \msp2 | \msp2
\msp2 \ulP _{\nu} \msp2 | \msp2 
\vec{e}\msp2 (\mathrm{T}_{\mu'}^{\msp2 0} ) \msp2 \rangle \Eq 0$ 
as expected for any set of $\nu$, $\mu$, and $\mu '$, and 
this confirms the stability of the equilibrium structure.~\cite{HT}  
This T$^0$-multiplet governs the three acoustic modes 
near $\bkk = \bm{0}$, and therefore  
their $\bkk$-dependence starts at second order, 
$\lambda_\alpha (\bkk ) \sim c_{\alpha}^2 \msp1 \bkk ^2$ ($\alpha$ = 1--3), 
where the coefficients are the sound velocities,  
$c_\alpha = \lim_{\bkk \to \bm{0}} \omega_{\alpha}(\bkk ) / |\bkk|$.  
One can prove that 
the two transverse modes share an identical 
velocity $c_1 = c_2 = c_\perp$, 
as in monoaxial chiral systems.~\cite{HT} 

Thus, linear split is possible only for the T$^{\pm}$-levels, 
and 
it is determined by 
three pseudoscalar quantities 
with time-reversal even parity, 
$\chi_{11}$, $\chi_{22}$, and $\chi_{12}$.
They are defined as 
\begin{equation}
\chi_{nn'} \equiv -\FRAC{i}{6} \sum_{\nu \mu \mu'} 
\varepsilon_{\nu \mu \mu'} \, 
\langle 
\vec{e} \msp2 (\mathrm{T}_{\mu}^{n} ) \msp2 | 
\msp2 \ulP _{\nu} \msp2 | 
\msp2 \vec{e} \msp2 (\mathrm{T}_{\mu'}^{n'} ) \rangle , 
\end{equation}
where $\varepsilon$ is the completely antisymmetric pseudotensor, 
and their values are 
$\chi_{11} \! = \!
 v_{xz}^{-} \msm2 + \msm2 4 \delta \msp2 v_{yz}^{+}$, 
$\chi_{22} \! = \!
 -v_{xy}^{-} \msm2 + 
 \msm2 v_{yz}^{-} \msm2 - 
 \msm2 4 \delta \msp2 v_{yz}^{+}$, 
and 
$\chi_{12} \! = \!
 \FRAC12 v_{yy}^{-} \msm2 - \msm2  2 \delta \msp2 
( v_{yy}^{+} \msm2 - \msm2 v_{zz}^{+})$. 
With these, the matrix elements necessary for degenerate perturbation 
calculations read as 
\begin{equation}
\langle \vec{e} \msp2 (\mathrm{T}_{\mu}^\sigma ) \msp2 | 
\msp2 \ulP _{\nu} \msp2 | \msp2
\vec{e} \msp2 (\mathrm{T}_{\mu'}^\sigma ) \rangle 
= 
- i \msp2 \gamma_{\sigma}^{} \msp2 
\varepsilon_{\nu \mu \mu '} 
= 
\gamma_{\sigma}^{} \msp2 
( \sfL _{\nu}^{} )_{\mu \mu'}^{} , 
\ \  (\sigma = \pm).  
\end{equation}
where 
\vspace{-8pt}
\begin{align}
\gamma_{\sigma}&= 
\FRAC12 (\chi_{11} + \chi_{22} )  + \sigma \msp2 
\mathrm{Re} \msp2 
\Bigl( e^{\msp1 i \theta} \, 
\Bigl[ \FRAC12 ( \chi_{11} - \chi_{22} )  +  
i \msp2 \chi_{12} \Bigr]  \, \Bigr) 
\nonumber\\
&=
\FRAC12 (v_{xz}^{-} + v_{yz}^{-} - v_{xy}^{-} ) 
+ \FRAC12 \msp2 \sigma \msp2 
\mathrm{Re} \msp2 
\bigl[ e^{\msp1 i \theta} 
( \alpha^{-} + 4 \delta \msp2 \alpha^{+} ) \bigr] , 
\end{align}
with 
$\alpha^{-} $=$ 
v_{xy}^{-} \msm2 + \msm2v_{xz}^{-} \msm2 - \msm2 v_{yz}^{-}
+ i \msp2 v_{yy}^{-}$ 
and 
$\alpha^{+} $=$ 
2 v_{yz}^{+} \msm2 - \msm2 i (v_{yy}^{+} \msm2 - \msm2 v_{zz}^{+})$.  
Note that $\sfL _\nu$'s are nothing but the angular momentum 
operators with spin 1.  
Thus, the reduced dynamical matrix has 
the $\bkk$-linear term in its expansion,  
$\ulD (\bkk ) \msp2 \bigr|_{\mathrm{T}^\sigma} \sim 
\lambda (\mathrm{T}^\sigma ) \mathsf{1} + 
\gamma_\sigma \msp2 \bkk \msp1 \cdot \msp1 \bLL
$,  
which has the form of spin-1 Weyl Hamiltonian.~\cite{ZhangT,MiaoH,LiH} 
The coefficient $\gamma _{\sigma}$ 
couples the polar ($\bkk$) and axial ($\bLL$) vectors, 
and one can see that $\gamma_\sigma$ inherits 
its time-reversal even pseudoscalar property from $\chi_{nn'}$'s.  
Reversing lattice handedness ($\delta \to -\delta$) 
interchanges $\sfv \supS$ and $\sfv \supL$.  
Thus, $\sfv ^-$ acquires a minus sign while $\sfv ^+$ remains unchanged, 
and this flips the sign of $\gamma _\sigma$.  
Recall that $\theta$ is determined from $\sfv ^+$ alone.  

Around $\bkk \Eq \bzero$, the energy dispersion of chiral phonons is 
\begin{equation}
\omega_{\sigma,h} (\bkk ) \msm2 \sim \msm2 
\omega_{\sigma}^{\scriptscriptstyle 0} + 
c_\sigma \msp2 h \msp2 |\bkk |, 
\ \ \ \mbox{where} \hspace{2mm}
c_\sigma \equiv \gamma_{\sigma} / (2 \omega_{\sigma}^{\scriptscriptstyle 0} ) , 
\end{equation}
for the T${}^{\sigma}$--mode.   
Here, 
$\omega_{\sigma}^{\scriptscriptstyle 0} =
 \lambda (\mathrm{T}^{\sigma})^{1/2} =
 [V_0 \PLUS \sigma \msp2 |\Delta (T)| \,]^{1/2}$.   
The quantum number $h \in \{ 0, \pm 1 \}$ is helicity, 
\textit{i.e.}, 
the projection of angular momentum 
$\underline{h} = \hat{\bkk} \cdot \underline{\bm{L}}$ 
for the momentum direction 
$\hat{\bkk} \Equiv \bkk / |\bkk |$, 
where $\underline{\bm{L}}$ is the angular momentum 
operator in the 12-dimensional full space.  
When $\bkk$ points to one of the trigonal-axis directions  
$\bbeta_i$, $h$ agrees with the CAM up to its sign. 
Not only that, 
the phonon eigenmode is also an asymptotic eigenvector 
of 
$\underline{h}$ for any $\bkk$ 
as $|\bkk | \to 0$, which we will discuss in 
more detail afterwards.  
This shows that the linear split is Dirac-cone type 
($h = \pm 1$) 
in addition to a flat mode ($h=0$).  
The velocity $c_\sigma$ differs between the $\mathrm{T}^+$- and 
$\mathrm{T}^{-}$-modes.  
Note that the operator 
$\ulbP$ 
has 
nonvanishing matrix elements between 
T${}^+$- and T${}^-$-modes. 
However, they have no contribution 
to linear split, and 
take effects from the order of $k^2$.  

Let us analyze the pseudoscalars 
$\chi_{11}$, $\chi_{22}$, and $\chi_{12}$
in more detail 
by relating them to 
the spatial configuration of stiffness tensors.  
Calculating the matrix elements, they read as 

\begin{align}
&\chi_{nn'}  = 
-\FRAC16 
\sum_{\mu \nu \nu'} 
\sum_{b \ne b'} 
\varepsilon_{\mu \nu \nu'} \, 
\Bigl[ \sfS{b}^{-n} \, 
\Bigl( \sfF _{\mu}^{(bb')} - \sfF _{\mu}^{(b'b)} \Bigr) \, 
\sfS{b'}^{n'} \Bigr]_{\nu \nu'}
\\[-4pt]
&\sfF _{\mu}^{(bb')} \equiv 
\sum_{p=S,L} 
( \btt _p^{(bb')})^{}_{\mu} \msp2 
\sfv _p^{(bb')} 
=
\bigl[ \sfR ^{(bb')} (-4 \delta \msp2 \baa _1 ) \bigr]^{}_{\mu} \, 
\sfR ^{(bb')} \circ \sfv ^{+} 
\nonumber\\[-14pt]
&\hspace{4cm}+ 
\bigl[ \sfR ^{(bb')} \baa _3 \bigr]^{}_{\mu} \, 
  \sfR ^{(bb')} \circ \sfv ^{-} .  
\end{align}
In deriving this, 
we have used the fact of 
$\btt \supS \msm2 + \msm2 \btt \supL$=$-8 \msp2 \delta \msp2 \baa _1$ 
and 
$\btt \supS \msm2 - \msm2 \btt \supL$=$ 2 \msp2 \baa _3$.  
One should note that each symmetry operation 
$\sfR ^{(bb')}$ comprises matrix elements 
restricted to the values 0 and $\pm 1$.  
Thus, reversing handedness 
($\delta \msm2 \to \msm2 -\delta$ and 
$\sfv \supS \msm2 \leftrightarrow \msm2 \sfv \supL$) 
changes the sign of $\btt \supS \msm2 + \msm2 \btt \supL$ and $\sfv ^{-}$, 
and this changes the global sign of 
$\sfF _{\mu}^{(bb')}$.  
The part of $\btt \supS \msm2 + \msm2 \btt \supL$ 
describes the structural part of the system's chirality, 
while $\sfv ^{-}$ does the part of spatial 
configuration of stress tensors.  
The two parts contribute additively to $\chi_{nn'}$'s.  

\begin{figure}[t]
\centering{
\includegraphics[width=8cm]{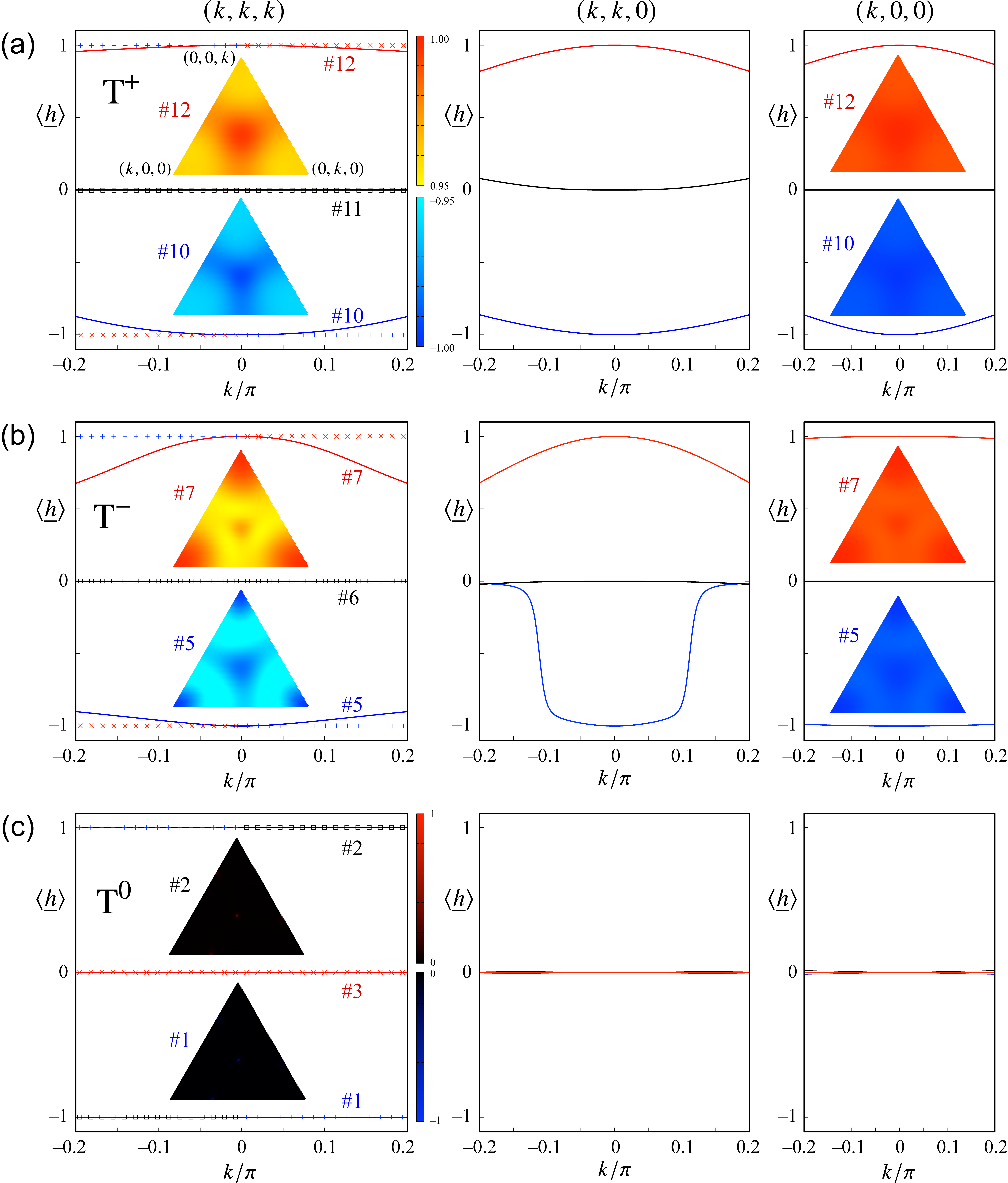}
}
\caption{
(Color online)
Helicity expectation value $\langle \underline{h} \rangle$ 
calculated for those states that are smoothly connected to 
the triplet levels at $|\bkk | = 0$.  
The numerals marked with $\#$-mark represent the mode number $\alpha$
arranged in ascending order of energy near $\bkk = \bm{0}$.  
Three states of the T$^{-}$ mode 
undergo multiple level crossings along $(k,k,k)$-direction 
in the shown region.  
The results are shown for those that are \textit{continuously} 
connected at those crossings.   
When $\bkk  = (k,k,k)$, CAM is always a good quantum number 
and its value $m$ is also shown by symbols.  
Insets: Color maps of the $\hat{\bkk}$-direction dependence 
of $\langle \underline{h} \rangle$ of specified phonon states, 
using an octant projection based on barycentric coordinates.   
The magnitude $|\bkk |$ is fixed at $0.1\pi$ in the left column 
and $0.05 \pi$ in the right column.  
The same color scale is used for 
the insets in (a) and (b).
}
\label{fig:helicity}
\end{figure}

Finally, we examine the effectiveness of 
the first-order expansion in $\bkk$.  
The phonon dispersion in Fig.~\ref{fig:disp} 
shows that linear energy split in the two triplets T$^{\msp2 \pm}$ 
quickly transforms into more complex $\bkk$-dependence 
as $|\bkk|$ increases.  
We can also examine this behavior by 
calculating helicity expectation value 
$\langle \underline{h} \rangle$ of each eigenvector. 
In the asymptotic region where 
the $\bkk$-linear term is dominant, 
all the eigenvectors $\vec{d}_{\alpha} (\bkk )$ of 
the triplets T$^{\msp1 \pm}$ are 
also approximate eigenvectors of 
$\underline{h} = \hat{\bkk} \cdot \underline{\bm{L}} $ and 
thus $\langle \underline{h} \rangle$ is close to 
either 0 or $\pm 1$.  
For this calculation, we need $\underline{\bm{L}}$ operator 
in the full space,
and may simply define this with 
the outer products of basis vectors as 
\begin{equation}
\underline{L}_{\, \nu} = 
-i \sum_{n=0,1,2} \ \sum_{\mu , \mu'} \, \varepsilon_{\nu \mu \mu'} \ 
\vec{e} \msp1 (T_{\mu}^{\msp1 n}) \, \, 
{}^t \vec{e} \msp1 (T_{\mu '}^{\msp1 n}) .  
\end{equation}

The calculated results $\langle \underline{h} \rangle$ are 
shown in Fig.~\ref{fig:helicity} for the two triplets T$^{\msp1 \pm}$ 
(optical modes) together with the T$^{\msp1 0}$-triplet 
(acoustic modes).  
For either of T$^{\msp1 \pm}$, the behavior of 
$\langle \underline{h} \rangle$ at $|\bkk | \to 0$ confirms 
our expectation and its value takes 
$-1$, 0, and 1 for the three states in the ascending order of energy.  
This holds for any direction of $\bkk$ 
as shown in the insets of the figure in the right column.
As $\bkk$ deviates from $\bm{0}$, $\langle \underline{h} \rangle$ 
changes continuously but the change is not so large 
until $|\bkk |$ becomes a few tenth of $\pi$.  
This change depends on the direction of $\bkk$ vector, 
since it is due to higher-order terms in the $\bkk$-expansion 
of $\ulD (\bkk )$. 
These higher-order contributions give rise to the 
small $\hat{\bkk}$-direction dependence 
shown for $|\bkk | = 0.1 \pi$ 
in the insets in the left column in Fig.~\ref{fig:helicity}.  
While this dependence vanishes as $|\bkk | \to 0$, 
it exhibits a three-fold rotation symmetry 
but lacks mirror symmetries, manifesting 
the chiral nature of the lattice.  
A noticeable change happens in the \#5 state
for $\bkk \msp1 $=$ \msp1 (k,k,0)$.  
This state undergoes a level repulsion around 
$k$$\msp1 \sim \msp1 $$\pm 0.1 \pi$ 
in that direction, and its character 
is no longer a triplet state beyond that point.  
The result for the acoustic mode T$^{\msp1 0}$ is 
interesting.  
As discussed before, the $\bkk$-linear spin-1 Weyl 
effective model does not apply to this part, 
but $\langle \underline{h} \rangle$ is quantized 
for $\bkk =(k,k,k)$, 
while almost vanishing if $\bkk$ tilts from 
this screw-axis direction even slightly.  
This is shown in Fig.~\ref{fig:helicity}(c).
The quantization for the $(k,k,k)$ direction 
originates from that fact that the CAM is a good 
quantum number.  
The three-fold rotation about the $\langle 111 \rangle$ axis,  
$\underline{C}_3$, is a symmetry operation, 
since this keeps the wavevector $\bkk =(k,k,k)$ invariant, 
and each dynamical matrix eigenvector 
is simultaneously 
an exact eigenvector of this rotation, 
$\underline{C}_{\msp1 3} \, \vec{d}_{\alpha} (\bkk ) 
= \exp (- i \msp1 m \msp1 \theta_3 ) \, \vec{d}_{\alpha} (\bkk )$ 
with $\theta_3 \equiv 2\pi/3$ and $m \in \{ 0, \pm 1 \}$.  
Because of a large energy gap between acoustic and optical 
modes, the acoustic mode eigenvectors are predominantly 
composed of the T$^{\msp1 0}$ bases, 
and in that case, 
$\underline{C}_{\msp1 3} \approx 
\exp [- i \msp1 \theta_3 \msp1 (
\underline{L}_{\msp1 x} \msm2 + \msm2
\underline{L}_{\msp1 y} \msm2 + \msm2 
\underline{L}_{\msp1 z} )/\sqrt3
] 
= 
\exp (\mp i \msp1 \theta_3 \msp1 \underline{h})$.  
Thus, the strict quantization of $m$ 
guarantees a nearly perfect quantization of 
helicity.  
This does not apply for the optical mode eigenvectors, 
which contain also considerable weights of 
A and E bases when $|\bkk |$ is not so small.  

In summary, we have investigated in this letter 
the phonon energy dispersion 
in a cubic chiral lattice 
and demonstrated that pseudoscalars 
determining the $k$-linear split of transverse optical modes 
are electric toroidal monopoles.  
These monopoles are defined from 
the spatial configuration of stiffness tensors 
together with the chirality parameter of structure.  

Many of our findings regarding the asymptotic 
behavior near $\bkk = \bm{0}$ 
arise from the system's symmetry 
and are applicable to general cubic chiral lattices.  
Specifically: 
(i) Acoustic modes exhibit an isotropic dispersion 
up to $k^1$-order, 
and their two transverse branches share the identical 
sound velocity regardless of polarization handedness.  
(ii) Singlet or doublet optical modes at $\bkk = \bm{0}$ 
exhibit energy dispersion whose $k$-dependence 
start at second or higher order.  
(iii) Triplet optical modes at $\bkk = \bm{0}$ 
exhibit linearly split dispersions according to 
helicity of each mode.  
This splitting is described by spin-1 Weyl Hamiltonian 
with a pseudoscalar coupling constant set by 
the system's electric toroidal monopoles.   
(iv) For 
$\bkk \parallel \langle 111 \rangle$, 
the phonon helicity asymptotically coincides with CAM up to a global sign, 
while the latter is a good quantum number 
irrespective of the value of $k$ for this direction. 

\vspace{6pt}
\begin{spacing}{0.8}
\begin{acknowledgment}
The authors gratefully acknowledge stimulating discussions 
with J. Kishine, Y. Kato, T. Satoh, Y. Togawa, and K. Hattori.
This work was supported by a Grant-in-Aid for Scientific Research 
(Grant No. JP23K03288) 
from the Japan Society for the Promotion of Science.
\end{acknowledgment}
\end{spacing}

\end{document}